\pdfoutput=1
\documentclass[runningheads]{llncs}
\usepackage{graphicx}
\usepackage{multirow}
\usepackage[symbol]{footmisc}
\usepackage[T1]{fontenc}

\begin{document}
\title{Combining Multi-Sequence and Synthetic Images for Improved Segmentation of Late Gadolinium Enhancement Cardiac MRI}
\titlerunning{Multi-sequence and synthetic images for LGE-MRI segmentation}
%
\author{V\'ictor M. Campello\inst{1}, Carlos Mart\'in-Isla\inst{1}, Cristian Izquierdo\inst{1}, Steffen E. Petersen\inst{3,4}, Miguel A. Gonz\'alez Ballester\inst{2,5} \and Karim Lekadir\inst{1}}
\authorrunning{V. M. Campello et al.}
%
\institute{Dept. Matem\`atiques i Inform\`atica, Universitat de Barcelona, Spain\\
\email{{victor.campello}@ub.edu}\\
\and BCN-MedTech, DTIC, Universitat Pompeu Fabra, Barcelona, Spain\\
\and Barts Heart Centre, Barts Health NHS Trust, London, United Kingdom \\
\and William Harvey Research Institute, NIHR Barts Biomedical Research Centre, Queen Mary University of London, United Kingdom \\
\and ICREA, Barcelona, Spain}
\maketitle              
\begin{abstract}

Accurate segmentation of the cardiac boundaries in late gadolinium enhancement magnetic resonance images (LGE-MRI) is a fundamental step for accurate quantification of scar tissue. However, while there are many solutions for automatic cardiac segmentation of cine images, the presence of scar tissue can make the correct delineation of the myocardium in LGE-MRI challenging even for human experts. As part of the Multi-Sequence Cardiac MR Segmentation Challenge, we propose a solution for LGE-MRI segmentation based on two components. First, a generative adversarial network is trained for the task of modality-to-modality translation between cine and LGE-MRI sequences to obtain extra synthetic images for both modalities. Second, a deep learning model is trained for segmentation with different combinations of original, augmented and synthetic sequences. Our results based on three magnetic resonance sequences (LGE, bSSFP and T2) from 45 different patients show that the multi-sequence model training integrating synthetic images and data augmentation improves in the segmentation over conventional training with real datasets. In conclusion, the accuracy of the segmentation of LGE-MRI images can be improved by using complementary information provided by non-contrast MRI sequences.

\end{abstract}
\section{Introduction}

Late gadolinium enhancement magnetic resonance imaging (LGE-MRI) is widely used to assess presence, location and extent of regional scar or fibrotic tissue in the myocardium. Whilst LGE-MRI is a well-established technique and key to many cardiovascular magnetic resonance (CMR) examinations there are challenges in quantification and interpretation due to a number of factors. Image analysis depends on image quality which can be affected by suboptimal CMR acquisition. Correct inversion times (TI) need to be identified and then TI require appropriate adjustments to allow good `nulling' of remote, unaffected myocardium. This ensures optimal contrast between scar/fibrosis (bright) and normal, remote myocardium (dark). Timing after contrast administration is important to allow not only sufficient wash-out of contrast agent (gadolinium chelate) from the remote myocardium but also from the blood pool. Images acquired too early will leave the blood pool bright which makes differentiating subendocardial infarct from blood pool challenging.

In the existing literature, two main families of techniques have been proposed to automatically segment LGE-MRI data. The first one segments directly the LGE-MRI images by using different techniques such as graph-cuts \cite{alba2014automatic}, atlas-based registration \cite{kurzendorfer2017fully}, or more recently Convolutional Neural Networks (CNNs) \cite{yue2019cardiac}. However, these techniques generally lack robustness due to the limited availability of LGE-MRI datasets for training. As a result, the second family of techniques has considered exploiting other cardiac MRI sequences to provide additional signals for guiding more robustly the segmentation process. For instance, some researchers \cite{wei2011myocardial,tao2015automated} proposed to segment first cine-MRI images and to propagate the obtained contours into the LGE-MRI images through image registration. Similarly but by using additional sequences, the authors in \cite{zhuang2018multivariate} implemented an atlas-based segmentation approach combining information from balanced-Steady State Free Processing (bSSFP), LGE and T2 sequences. However, these techniques are highly dependent on the image registration step, which is challenging due to the inherent differences between the cardiac MRI sequences. 

In addition, in order to improve segmentation and increase the model robustness over unseen data, image synthesis has been proposed recently. The most common model combines generative adversarial networks (GANs) with a cycle-consistency constrain for image-to-image translation and two segmentation networks, one for each image domain, trained end-to-end in order to benefit from a combined loss function. This model has been applied for cross-modality segmentation improvement \cite{zhang2018translating,cai2019towards}, domain adaptation across scanners \cite{cai2019towards} or across modalities \cite{chen2019synergistic} and segmentation of an unlabeled target modality using only the source ground truth \cite{huo2018synseg,zhang2018task}. Alternatively, a GAN can be trained to generate synthetic images from masks according to some conditional value, like the dataset style, as in the case of retinal fundus images for vessel segmentation \cite{zhao2018supervised}.

In this paper, we propose an approach to circumvent the need for image registration, while addressing the lack of LGE-MRI images for training. Concretely, we implement a CNN-based approach that is capable of learning key properties of the cardiac structures simultaneously from multiple cardiac MRI sequences. Furthermore, image synthesis and data augmentation are used to generate new examples that take into account both the global appearance of LGE-MRI data and the local appearance of scar tissues. With this approach, direct deep learning based segmentation of LGE-MRI is enabled without the need for inter-sequence image registration and while exploiting the richness of multi-sequence cardiac MRI.

\section{Method}

\subsection{Dataset}
\subsubsection{Data description} The LGE-MRI dataset used in this paper was provided as part of the Multi-Sequence Cardiac Magnetic Resonance Segmentation Challenge (MS-CMRSeg). It consists of 45 patients from Shanghai Renji Hospital that were scanned using three MRI sequences: bSSFP, LGE and T2. Ground truth segmentations of the left ventricle (LV), right ventricle (RV) and myocardium (MYO) were provided for some of the cases according to the distribution in Table~\ref{tab1} (second row). Even though all sequences were acquired and selected for the end-diastolic cardiac phase, there were differences in the shape of the cardiac boundaries consistently between the three sequences for the same patient. Moreover, the slices were not aligned between the sequences in the direction of the ventricular axis, which further complicates the application of image registration. Note that all patients in the sample suffer from cardiomyopathies and that every LGE-MRI image presents a scar of variable size within the myocardial wall.

\begin{table}
\centering
\caption{MS-CMRSeg sequences details.}\label{tab1}
\begin{tabular}{l|c|c|c|}
	 &  bSSFP & LGE & T2 \\
	\hline \hline
	Number of patients & 45 & 45 & 45 \\
	Segmented patients & 35 & 5 & 35 \\
	Number of slices & 8 -- 12 & 10 -- 18 & 3 -- 7 \\
	Slice thickness ($mm$) & 8 -- 13 & 5 & 12 -- 20 \\
	TR/TE ($ms$) & 2.7/1.4 & 3.6/1.8 & 2000/90 \\
	In-plane resolution ($mm$) & $1.25\times 1.25$ & $0.75 \times 0.75$ & $1.35\times 1.35$ \\
	\hline
\end{tabular}
\end{table}

\subsubsection{Data pre-processing} As a first step, intensity bias correction was applied to all sequences to correct for potential artifacts and the intensity histograms of all images were matched to a common one to obtain coherent appearances across images. Furthermore, before the training process, all images were interpolated and cropped so that they had a pixel size of $256\times256$ and the same resolution. They were also normalised such that the mean intensity and the standard deviation equal $0.5$, thus ensuring most of the input values to be positive in between 0 and 1 for convenience in later representation of the images.

\subsection{Increasing training sample}
Before describing the CNN model implemented in this paper for LGE-MRI segmentation, this section presents two methods used to increase the number of training data and obtain higher LGE-MRI variability.

\subsubsection{Data augmentation} 
\begin{figure}
\centering
\includegraphics[trim={0 3cm 0 2cm}, clip, width=\textwidth]{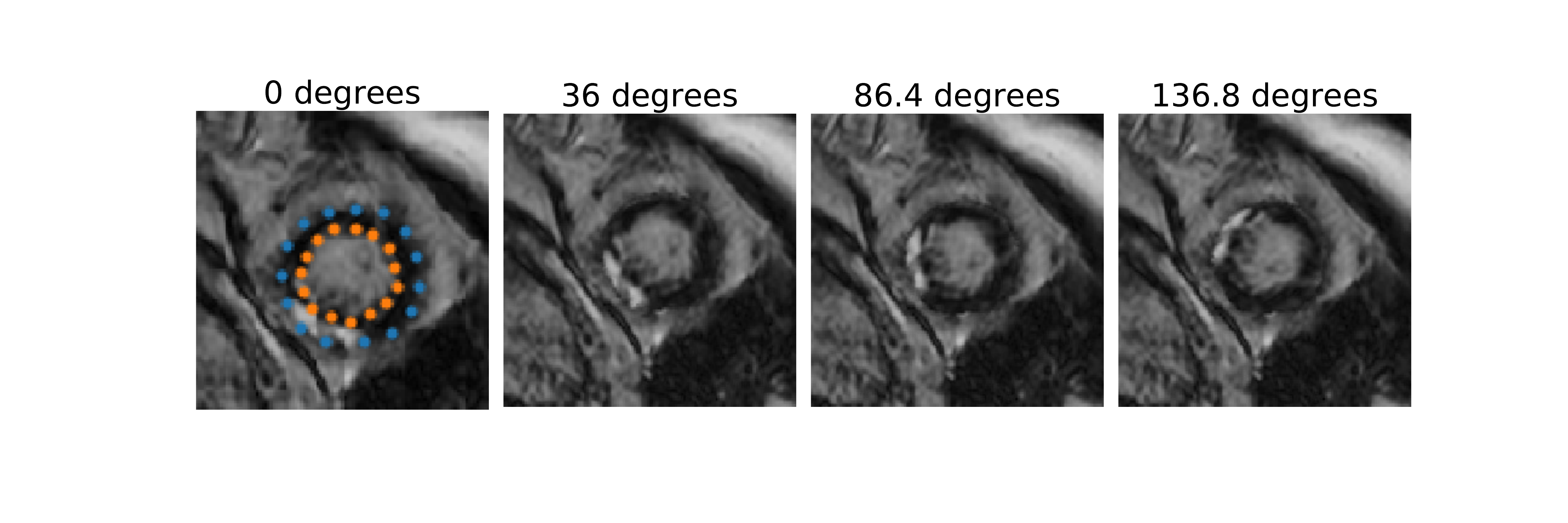}
\caption{Example of three rotations of the myocardial wall with respect to the whole image by using the landmarks provided in the leftmost image. This shows the changes in the location of the scar tissues} \label{fig:rotation}
\end{figure}

By using the provided segmentations, a set of 50 landmarks were evenly placed around the epicardium and endocardium. With these, the myocardium and left ventricle were rotated relative to the rest of the image, as shown in the examples in Figure~\ref{fig:rotation}, in order to obtain an augmented dataset with varying locations of the scar tissues. Since the contour of the epicardium is not perfectly round in general, a Gaussian filter of size $3\times 3$ was applied around the outer boundary to smooth the transition between the rotated and fixed regions, thus preventing image intensity discontinuities. A total of twenty $7.2$ degrees rotations were applied. Thus, the LGE-MRI dataset was multiplied by a factor of 20 and the location of the scar in the myocardium ranged between the initial position and 144 degrees clockwise. This augmentation technique increases the variability in the scar locations within the myocardial wall that was otherwise very low due to the small number of patients available for training. Furthermore, further data augmentations were obtained by applying small rotations of the input images up to 15 degrees before training.

\subsubsection{Image synthesis} The rationale behind the proposed image synthesis is that there are many more segmented cine-MRI datasets available open-access or in clinical registries for training CNN models. Thus, to increase the number of annotated LGE-MRI cases for training, image synthesis from cine-MRI images sequences is proposed. To achieve this, the CycleGAN method \cite{zhu2017unpaired} was implemented using the PyTorch library provided at this link\footnote[2]{https://github.com/junyanz/pytorch-CycleGAN-and-pix2pix}.

This method translates images from one domain to another without the need for image registration or for the sequences to be from the same patients. It consists of a pair of generators $G_{LGE}$, $G_{bSSFP}$ and a pair of discriminators $D_{LGE}$, $D_{bSSFP}$ that have opposed goals. The generator $G_{LGE}$ ($G_{bSSFP}$) transforms the bSSFP (resp. LGE) sequence into a realistic LGE (bSSFP) image, while the discriminator $D_{LGE}$ ($D_{bSSFP}$) attempts to distinguish between real and fake LGE (bSSFP) sequences. To achieve a good image translation between the two sequences, the loss function contains two terms: (1) an adversarial loss for each target domain that accounts for the similarity between the generated and real images, and (2) a cycle consistency loss that ensures that the transformed image $G_{LGE}(X)$ ($G_{bSSFP}(Y)$) is transformed back to $X$ ($Y$) through $G_{bSSFP}$ ($G_{LGE}$).

\begin{figure}
\includegraphics[trim={0 3cm 0 1cm}, clip, width=0.9\textwidth]{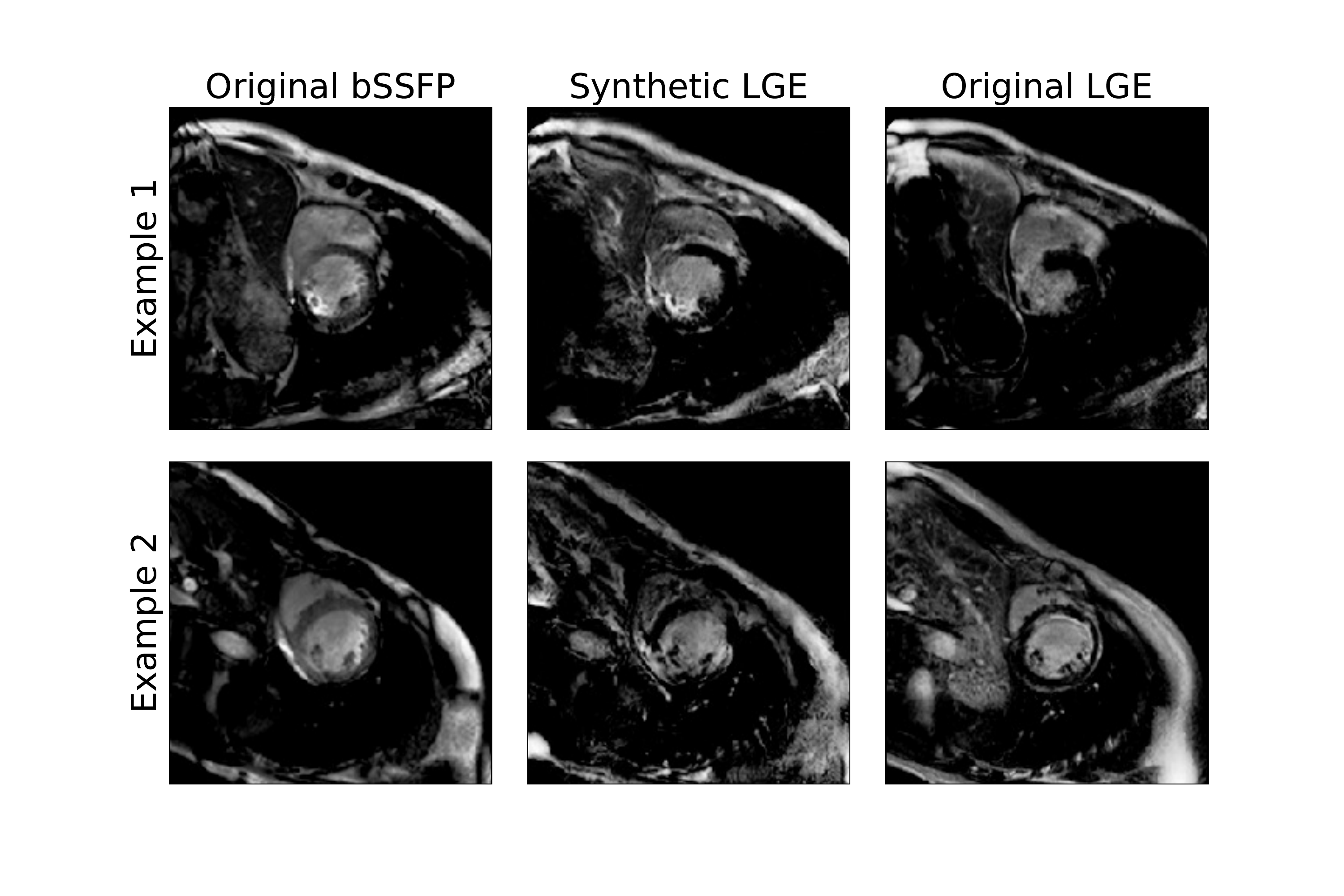}
\caption{Examples of synthetic LGE-MRI images. The leftmost column are the original cine images, the central column shows the transformed images to the LGE domain and the rightmost column is the most similar slice from the real LGE sequences, since they were not registered/aligned.} \label{fig:synthesis}
\end{figure}

For the training of the CycleGAN model, all slices from the 45 patients for the LGE and bSSFP sequences were used during 200 epochs. The training took 12 hours on a NVIDIA 1080 GPU with a batch size of $1$. The Adam optimizer was used with learning rate of $2\times 10^{-4}$, with first and second moment decay rates of $0.5$ and $0.999$, respectively. Some examples for the generated images are shown in Figure~\ref{fig:synthesis}.

In order to evaluate the quality of the generated images, two segmentation models (like the one described in the next subsection) were trained using the bSSFP images and the synthetic LGE images separately. The obtained results are presented in Table~\ref{tab:quality}. In particular, the synthetic LGE images, that are anatomically similar to the original bSSFP, provide more information for the task of LGE segmentation.

\begin{table}
\centering
\caption{Average and standard deviation for the Dice score of segmentation results over the five labeled LGE volumes.}\label{tab:quality}
\begin{tabular}{l|c|c|c|c|c|c|}
	 & \multicolumn{2}{|c|}{LV} & \multicolumn{2}{|c|}{MYO} & \multicolumn{2}{|c|}{RV}   \\
	 \cline{2-7}
	 &  avg. & std. & avg. & std. & avg. & std. \\
	\hline \hline
	model trained w. bSSFP & \,0.503\, & \,0.406\, & \,0.370\, & \,0.301\, & \,0.515\, & \,0.434\, \\
	model trained w. synthetic LGE & 0.809 & 0.116 & 0.688 & 0.145 & 0.820 & 0.065 \\
	\hline
\end{tabular}
\end{table}

\subsection{CNN-based LGE segmentation} 
\begin{figure}
\includegraphics[width=\textwidth]{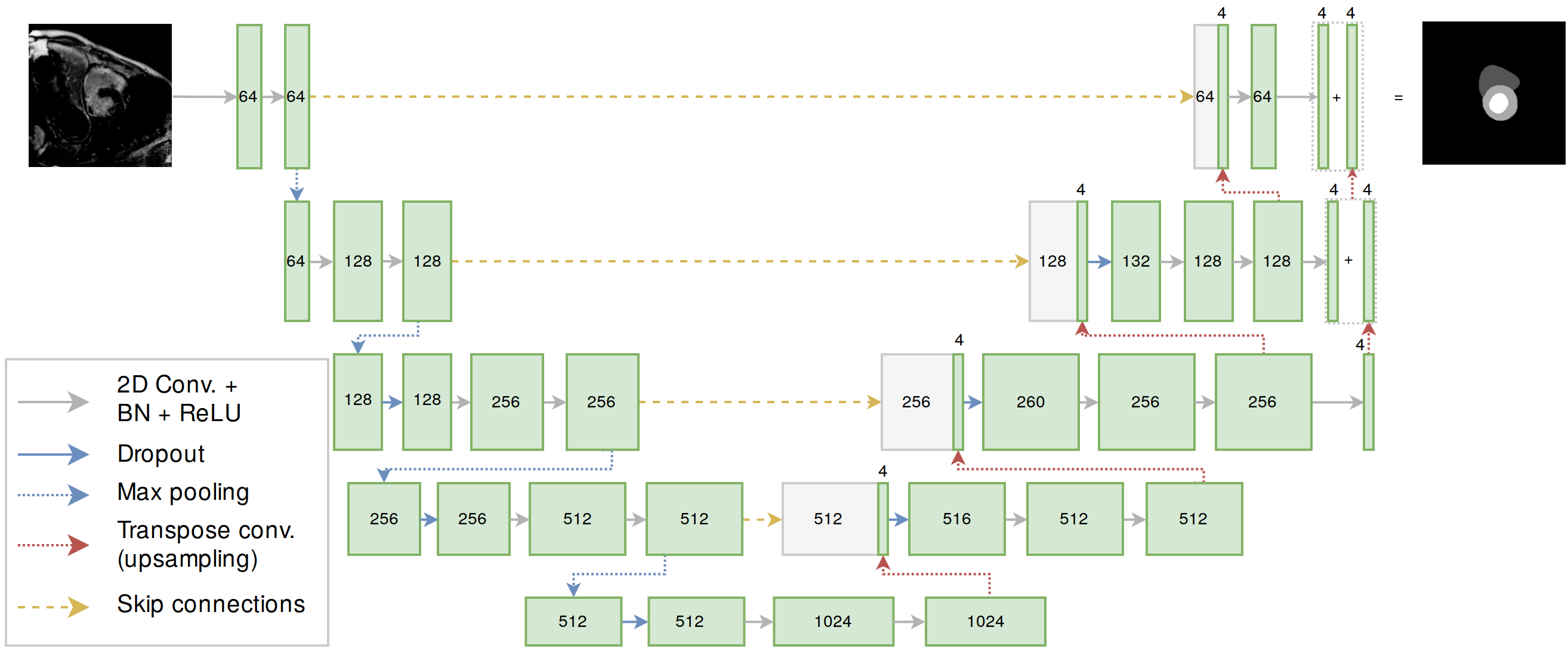}
\caption{Detailed architecture of the CNN model used for LGE segmentation. The numbers in the boxes correspond to the number of channels. Convolution operations have a kernel size of $3\times 3$ and stride of $1$, while transpose convolutions have a kernel size of $4\times 4$ and stride of $2$.} \label{fig:model}
\end{figure}

Once a large set of training sample was obtained from the original, augmented and synthetic images, a modified U-Net architecture \cite{ronneberger2015u} was used for the image segmentation by integrating two techniques: (1) a deep supervision term in the upsampling path as proposed in \cite{isensee2017automatic} that will act as lower-resolution masks that are convolved to condition the final predictions; and (2) a reduction of the number of filters after each upsampling operation to match the number of labels as proposed by \cite{baumgartner2017exploration}. Each image in the dataset was provided as a single channel input, thus forcing the model to differentiate between sequences with a unique set of weights. Additionally, in order to avoid overfitting given the sample size, dropout was used after every max pooling and upsampling operations, except for the high level features in the architecture, as shown in Figure~\ref{fig:model}.

During training, 20\% of the patients for each dataset was reserved for validation and early stopping. With a batch size of $8$ images, this model took less than 36 hours to achieve the best accuracy on the validation set after almost 90 epochs on a NVIDIA TITAN X GPU. The Adam optimizer was used with a learning rate of $10^{-4}$, with first and second moment decay rates equal to $0.9$ and $0.99$, respectively.

\section{Results}

In order to define the best trained CNN model for LGE-MRI segmentation, various training sets were used by varying the input sequences and combinations of image synthesis and scar augmentation, as follows:
\begin{enumerate}
    \item LGE sequences only;
    \item LGE and bSSFP sequences;
    \item All sequences (LGE, bSSFP and T2);
    \item All sequences plus MYO and LV rotations in LGE sequences;
    \item Number 1 plus synthetic LGE sequences;
    \item Number 2 plus synthetic LGE sequences;
    \item Number 3 plus synthetic LGE sequences;
    \item Number 4 plus synthetic LGE sequences.
\end{enumerate}

When evaluated on the validation set, the training set number 8 resulted in the best segmentations, showing the added value of image synthesis and data augmentation for LGE-MRI segmentation. Thus, we applied the corresponding CNN model to the test dataset composed of 40 LGE-MRI cases. The obtained segmentations were sent to the organizers of MS-CMRSeg Challenge for evaluation. The obtained results are summarized in Table~\ref{tab:testresults}, showing average dice scores of 90\% (LV), 87\% (MYO) and 81\% (RV). 

\begin{table}
\centering
\caption{Average and standard deviation for results over the test set.}\label{tab:testresults}
\begin{tabular}{l|c|c|c|c|c|c|}
	 & \multicolumn{2}{|c|}{LV} & \multicolumn{2}{|c|}{MYO} & \multicolumn{2}{|c|}{RV}   \\
	 \cline{2-7}
	 &  avg. & std. & avg. & std. & avg. & std. \\
	\hline \hline
	Dice score & \,0.898\, & \,0.045\, & \,0.810\, & \,0.061\, & \,0.866\, & \,0.051\, \\
	Jaccard index & 0.817 & 0.072 & 0.685 & 0.084 & 0.768 & 0.078 \\
	Surface distance ($mm$) & 2.0 & 0.8 & 1.8 & 0.5 & 2.3 & 0.9 \\
	Hausdorff distance ($mm$) & 11 & 4 & 12 & 4 & 16 & 7 \\
	\hline
\end{tabular}
\end{table}

Two remarks are important to note regarding the results reported in Table~\ref{tab:testresults}: (1) Despite the high variability in the LGE-MRI datasets, especially in the presence, extent and location of the scar tissues, relatively consistent results are obtained with standard deviations for the dice scores around 5\%. (2) Despite the availability of only 5 LGE-MRI volumes for training, the proposed approach was able to achieve comparable results to very recent deep learning techniques, which reported dice scores of $0.915\pm 0.052$ (LV), $0.812\pm 0.105$ (MYO) and $0.882\pm 0.084$ (RV) based on 5 times more training cases (25 LGE-MRI images). 
\cite{yue2019cardiac}. This indicates the value of the proposed inter-sequence synthesis and scar augmentation for generating richer training samples.

\begin{figure}
\vspace{-5mm}
\includegraphics[trim={0 3cm 0 1cm}, clip, width=\textwidth]{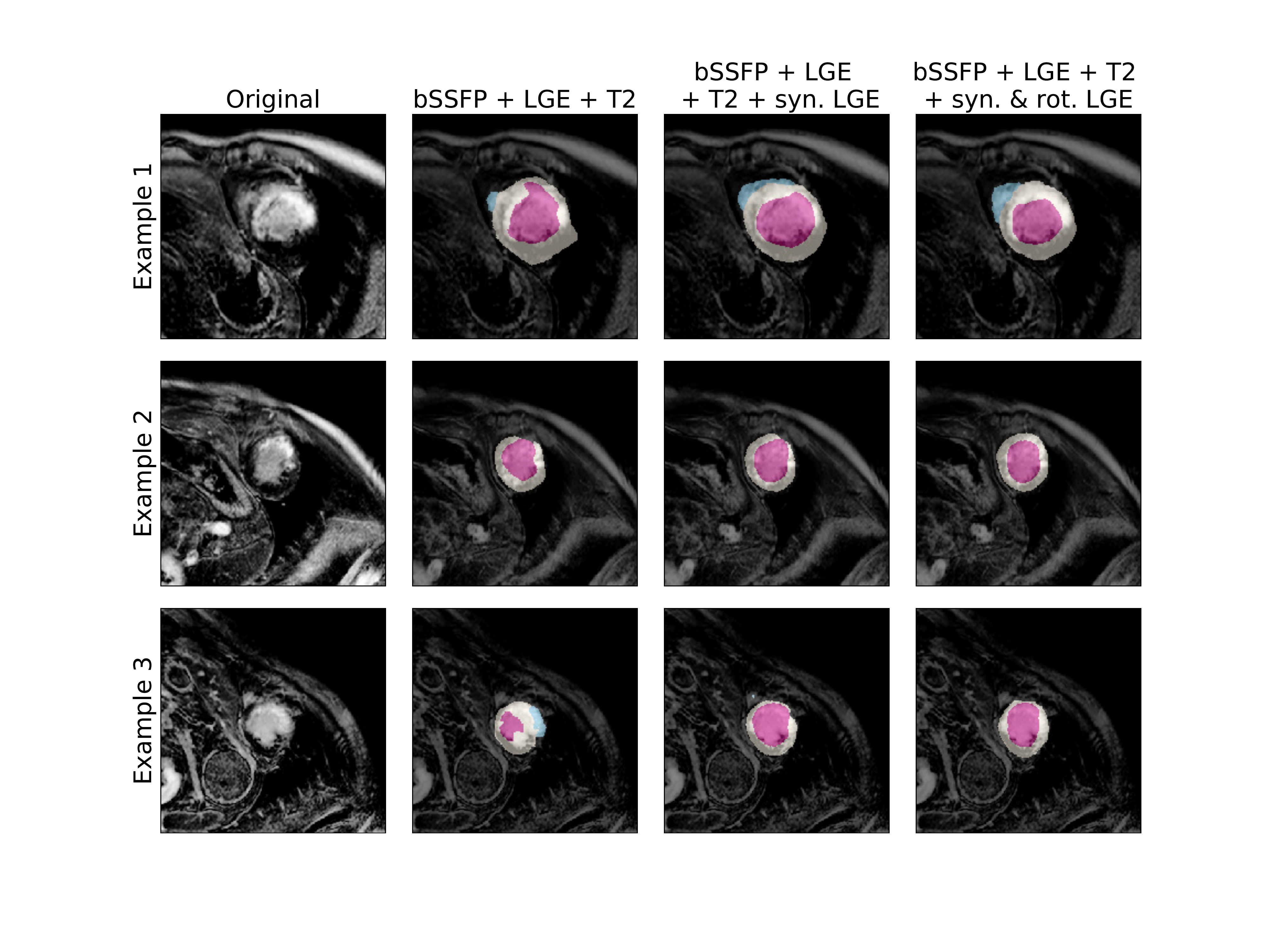}
\caption{Three segmentation examples as obtained by using different training combinations, showing the improvement achieved by integrating inter-sequence image synthesis (column 3) and scar tissue augmentation (column 4) during training.} \label{fig:segmentations}
\vspace{-5mm}
\end{figure}

Finally, for visual illustration, Figure~\ref{fig:segmentations} shows three segmentation examples as obtained in this study. Model number 3 (second column) introduces errors that are corrected when adding synthetic images (model number 7 in the third column). The last column shows that the segmentation further improves when integrating the scar tissue augmentation as proposed in this paper (model 8).

\section{Conclusions}

This paper proposed to address the limited availability of training samples for LGE-MRI segmentation by enriching the CNN models using two complimentary methods. Firstly, since samples of annotated cine-MRI sequences are more commonly available, image synthesis of LGE-MRI images was implemented using a CycleGAN approach, thus obtaining a larger number of LGE-MRI cases during training. Secondly, we performed LGE-specific data augmentation through shape-guided rotations of the myocardium, which increases the variability related to the location of the scar tissues in the myocardium. The validation shows consistent results across the datasets, indicating the potential of this approach for enhancing the richness and generalization of LGE-specific CNNs. 

Future work include the extension of the image synthesis to take into account local cardiac motion abnormality for synthesizing scar tissue, as well as the use of elastic deformations of the myocardium and scar to augment non-rigidly the LGE-MRI examples. Furthermore, extensive validation will be performed to assess in detail the relative importance of the different steps and sequences (bSSFP, T2) in enriching the CNN models for LGE segmentation.
\section{Acknowledgements}

This work was partly funded by the European Union's Horizon 2020 research and innovation programme under grant agreement No 825903 (euCanSHare project). SEP acts as a paid consultant to Circle Cardiovascular Imaging Inc., Calgary, Canada and Servier. SEP acknowledges support from the National Institute for Health Research (NIHR) Cardiovascular Biomedical Research Centre at Barts, from the SmartHeart EPSRC programme grant (EP/P001009/1) and the London Medical Imaging and AI Centre for Value-Based Healthcare. SEP and KL acknowledge support from the CAP-AI programme, London's first AI enabling programme focused on stimulating growth in the capital's AI Sector.

%
%
%
\bibliographystyle{splncs04}
\bibliography{bibliography.bib}

\end{document}